\documentclass{article}
\usepackage{epsf}
\usepackage{amssymb,amsmath}
\usepackage{color}
\newcommand{\revis}[1]{\textcolor{black}{#1}}

\begin{document}
\title{Elastic Lennard-Jones Polymers Meet Clusters -- Differences and Similarities}
\author{Stefan Schnabel$^1$\thanks{Corresponding author: stefanschnabel@physast.uga.edu}, Michael Bachmann$^2$\thanks{m.bachmann@fz-juelich.de}, and Wolfhard Janke$^3$\thanks{janke@itp.uni-leipzig.de}\\[.5cm]
{\normalsize{$^1$Center of Simulational Physics, University of Georgia}}\\
{\normalsize{$^2$Institut f\"ur Festk\"orperforschung, Theorie II, Forschungszentrum J\"ulich}}\\
{\normalsize{$^3$Institut f\"ur Theoretische Physik, Universit\"at Leipzig}}}
%
%
%
\date{}

\maketitle

\begin{abstract}
\noindent
We investigate solid-solid and solid-liquid transitions of elastic flexible off-lattice polymers with Lennard-Jones monomer-monomer interaction and anharmonic springs by means of sophisticated variants of multicanonical Monte Carlo methods. We find that the low-temperature behavior depends strongly and non-monotonically on the system size and exhibits broad similarities to unbound atomic clusters. Particular emphasis is dedicated to the classification of icosahedral and non-icosahedral low-energy polymer morphologies.
\medskip\\
\textit{Keywords:} Polymer crystallization, Mackay layer,
Lennard-Jones cluster, Conformational transition, Monte Carlo computer
simulation
\medskip\\
PACS: 05.10.-a, 36.40.Ei, 87.15.A-
\end{abstract}
\section{Introduction}

\revis{The understanding of the crystallization behavior of single flexible and elastic polymer chains is of particular interest in natural and technological molecular building-block systems, where functionalization is based on the individual structural properties of small molecules forming larger-scale composites. Among these are, e.g., hybrid systems of solid substrates and "soft" molecules, where it is known that the formation of regular structures can be mediated by the interaction with the amorphous or crystalline surface of the substrate \cite{X1}. Tuning the self-assembly of individual polymers near substrates is relevant for molecular nanoelectronic applications and materials design which has become a huge field of applied research. As an example from life sciences, capsomer proteins of spherical viruses assemble to form a perfectly icosahedral capsid hull, which shelters the viral genome \cite{X2}. Thus, frozen structures of single polymer chains can serve as the basic elements of larger assemblies on nanoscopic scales. As we will show here, the crystallization behavior of elastic polymers exhibits strong similarities to the cluster formation of colloidal (or atomic) particles.}

The general behavior of a polymer in solvent has already been subject of numerous studies \cite{KremerBinder,BinderMilchev}. At high temperatures or in good solvent the interactions of different parts of the polymer are of little relevance and due to entropic effects the dominating structures are extended random coils. \revis{This changes at the tricritical $\Theta$-point where the polymer behaves in three dimensions effektively like a Gaussian chain, up to logarithmic corrections \cite{Hager}.} The effective repulsion caused by excluded-volume effects and the monomer-monomer attraction are in perfect balance. In the globular regime, the chain is collapsed and compact conformations dominate which, having almost no internal structure, resemble a liquid state. At lower temperature, energy optimization leads to a freezing towards crystalline structures. In the thermodynamic limit, this transition is expected to be of first order, whereas for the $\Theta$-collapse a second-order phase transition is predicted \cite{Lifshitz,Khokhlov}.

Under certain conditions, the two transitions also may coincide in the thermodynamic limit as it was shown in a recent investigation of a bond-fluctuation model \cite{Rampf}. However, this is not a generic feature of flexible \cite{Thomas} and elastic polymers \cite{letter}.

While the coil-globule transition is well analyzed and further progress in the analysis of logarithmic corrections to the scaling behavior would require the investigation of \revis{extremely} long chains, less is known for the freezing transition especially in the case of off-lattice polymers. Due to the intrinsic similarities of the applied model to colloidal clusters it is useful to interpret the crystallization of elastic polymers with reference to atomic Lennard-Jones clusters and to use the corresponding nomenclature. In fact, if one replaces `atoms' by `monomers' many statements from the cluster field remain also true for polymers. Besides the numerous studies that have been performed on the freezing transition and ground-state properties of Lennard-Jones clusters, there are only a few publications on Lennard-Jones homopolymers. The authors of Ref.\ \cite{calvo} were able to show principal relationships between both classes of systems and motivated the transfer of terms and concepts from the cluster case to the field of polymers. Recent studies \cite{Liang,Australien1,Australien2,Seaton,Seaton2} mainly focused on the general behavior of elastic polymers \revis{but} did not refer in detail to the polymer length dependence of the freezing transition, the ground-state conformations or solid-solid transitions.

The aim of this work is to close this gap and to compare the low-temperature properties of polymers and clusters. The rest of the paper is organized as follows. In the subsequent section we will introduce the model and mention some problems that arise in the simulations. We then give a short review of the thermodynamic properties of Lennard-Jones clusters in Sec.~III, followed by the introduction of a geometrical order parameter in Sec.~IV. After the description of the method and the explanation of a few technical details in Sec.~V, we will present and discuss our results in Sec.~VI. Finally, the main findings of the paper are briefly summarized in the concluding part.

\section{Model}

We employ a model for elastic flexible polymers, where all monomers interact pairwisely via a truncated and shifted Lennard-Jones (LJ) potential:
\begin{equation}
	V_{\rm LJ}^{\rm mod}(r_{ij})=V_{\rm LJ}(\min(r_{ij},r_c))-V_{\rm LJ}(r_c).\\
\end{equation}
Here, $r_{ij}$ denotes the distance between the $i$th and $j$th monomer, $r_c$ is the cutoff distance, and
\begin{equation}
	V_{\rm LJ}(r)=4\epsilon[(\sigma/r)^{12}-(\sigma/r)^{6}] \\
\end{equation}
the standard LJ potential. In the following, $\epsilon$ is set to $1$ and $\sigma=r_0/2^{1/6}$ with the minimum-potential distance $r_0=0.7$. In this work we used and compared three different cutoff distances $r_c=2.5\sigma, 5\sigma, \infty$.

Covalent bonds are modeled by an additional FENE-potential \cite{FENE} for adjacent monomers:
\begin{equation}
V_{\rm FENE}(r_{ii+1})=-\frac{K}{2}R^2\ln\{1-[(r_{ii+1}-r_0)/R]^2\}.
\end{equation}
The potential possesses a minimum coinciding with $r_0$ and diverges for $r\rightarrow r_0\pm R$ with $R=0.3$ here. $K$ is a spring constant set to $40$.

Eventually, the polymer energy is given by:
\begin{equation}
E=\frac{1}{2}\sum_{ ^{i,j=1}_{i\neq j}}^N V_{\rm LJ}^{\rm mod}(r_{ij})+\sum_{i=1}^{N-1} V_{\rm FENE}(r_{ii+1}).
\end{equation}

\section{A Short Review of Lennard-Jones Clusters}

\begin{figure}
\centerline{\epsfxsize=4cm \epsfbox{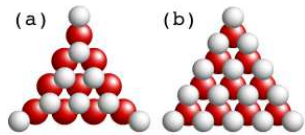}}
\caption{\label{fig:fig_1} (Color) Overlayer (bright) types on an icosahedron's face (dark): (a) anti-Mackay (hcp), (b) Mackay (fcc) \cite{Northby}.}
\end{figure}

Much work has been done in LJ cluster studies to identify and classify global energy minima. At least for $N\le200$ the ground states are known with high certitude \cite{Northby,Barron,basin_hopping}. Most global minima are constructed via a common scheme: usually an icosahedral core is covered by an incomplete overlayer, which gets more populated with increasing system size $N$ until a magic number ($N=13,55,147,309,561,923$ \cite{Mackay}) is reached and the next complete icosahedron is formed. Two types of overlayers occur: If only a few monomers are added to the core they gather at one of the faces filling the available space most efficiently. This leads to an hcp-like or anti-Mackay packing [Fig.~\ref{fig:fig_1}(a)]. With increasing system size, other faces are occupied and additional atoms adsorb at the corners. Thereby atoms on different faces interact only weakly. Hence, a different type of structure is energetically favored when a certain system size is exceeded. The external monomers adapt the partly fcc-like structure of the core and build a Mackay overlayer, which also allows the occupation of the core edges at the expense of non-optimal distances [Fig.~\ref{fig:fig_1}(b)]. With a few exceptions, one finds anti-Mackay ground states for clusters with $N=$14--30,56--81,85, whereas sizes $N=$31--54,82--84,86--146 lead to Mackay global energy minimum (GEM) \cite{Northby}.  For some systems, whose sizes differ notably from the optimal values $N=13,55,147,...$, non-icosahedral geometries become energetically competitive. The ground state of the 38-atom cluster is an fcc-like truncated octahedron \cite{Pillardy}, systems of 75-77 \cite{Doye1} and 102-104 \cite{Doye2} atoms form Marks decahedra  with a central fivefold symmetry axis and the cluster with 98 atom forms a Leary tetrahedron \cite{tetra98}.

When considering non-zero temperatures one finds that below the melting temperature clusters occupying Mackay-like conformations undergo a solid-solid transition where the overlayer transforms to anti-Mackay type \cite{LJcluster_heatcap}. Thereby the transition temperature increases with system size. Clusters with non-icosahedral GEM typically change at very low temperatures to icosahedral structures with larger population at intermediate energies.
\section{Geometrical Classification of Solid Elastic Polymers}
Structural changes between different geometries appear to be a characteristic feature of the cluster's behavior at low temperatures. Since we expect a similar behavior also for elastic polymers, we are interested in classifying the geometrical states of a conformation in order to specify conformational transitions. While many studies refer to bond orientation parameters \cite{bond_orient_para} that are calculated via spherical harmonics, we restrict ourselves to informations provided by the contact map which can be updated instantly during the simulation without further computational expenses. We consider two monomers as being in contact if their distance is smaller than a threshold $r_{\rm contact}$. Due to this definition also bonded monomers do not need to be in contact.
\begin{figure}
\centerline{\epsfxsize=8cm \epsfbox{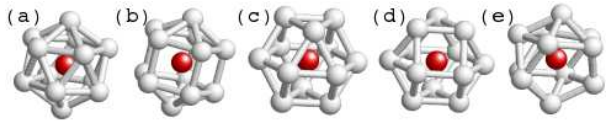}}
\caption{\label{fig:fig_2} (Color) Possible shell conformations ($r_c=2.5\sigma$) of a monomer possessing 12~(a-d) or 11~(e) neighbors. (a) icosahedron, (b) elongated pentagonal pyramid \cite{Johnson}, (c) cuboctahedron (fcc), (d) triangular orthobicupola \cite{Johnson} (hcp), (e) incomplete icosahedron. Sticks illustrate shell contacts, not bonds.}
\end{figure}
The total number of monomer contacts is not an appropriate measure for classification since in the interior of a frozen polymer every monomer has usually exactly 12 neighbors. Instead, the contacts between that 12 neighbors reflect their arrangement which corresponds to the local geometry. In Fig.~\ref{fig:fig_2}(a-d), different conformations of a monomer and its 12 neighbors are shown. Counting the contacts between the neighbors is a simple but efficient way to characterize different types: Only the icosahedral cell [Fig.~\ref{fig:fig_2}(a)] reveals 30 shell contacts corresponding to the 30 edges of an icosahedron. It always appears in the center of an icosahedral conformation. Consequently, if there is no such basic element, the global geometry cannot be icosahedral! On the other hand, icosahedral cells are also formed by a sufficiently large anti-Mackay overlayer at the corners of the icosahedral core. If the number of outer monomers is too small, one might find instead the defected icosahedral cell [Fig.~\ref{fig:fig_2}(e)], a monomer with 11 neighbors forming 25 shell contacts. The total number of both structures $n_{\rm ic}$ is a suitable \revis{``order''} parameter which allows \revis{a classification of the global geometry at low temperatures, given roughly by
\begin{equation}
n_{\rm ic}\qquad \left\{\begin{array}{l}
=0\qquad \mbox{non-icosahedral}\\
=1\qquad \mbox{icosahedral + Mackay,}\\
\ge2\qquad \mbox{icosahedral + anti-Mackay.}
\end{array} \right.
\label{p_n_ic}
\end{equation}
More precisely, if $n_{\rm ic}=0$, the polymer forms a non-icosahedral structure, e.g., it is decahedral or fcc-like; $n_{\rm ic}=1$ indicates icosahedral geometry with Mackay overlayer or a complete icosahedron which might possess a few monomers bound in anti-Mackay type. Finally, for $n_{\rm ic}\ge2$, the monomers form an icosahedral core with a considerably extended anti-Mackay overlayer.
The probabilities $p_{n_{\rm ic}}(T)$ for the different values of $n_{\rm ic}$ as a function of temperature provide the necessary information to reveal structural transitions.}

Figure~2(b) shows the elongated pentagonal pyramid with 25 shell contacts which is the basic module of five-fold symmetry axes in icosahedra and decahedra. It also occurs along the edges of an icosahedral core, which is covered by an anti-Mackay overlayer. Hence, it appears in icosahedral conformations with $N\ge 31$ and decahedral structures. Besides, it is formed at the edges of the central tetrahedron in conformations with a tetrahedral symmetry. An example is the ground state of the cluster with $N=98$ or low-energy minima for $N=159$ and $N=234$. In consequence, the total number of elongated pentagonal pyramids $n_{\rm epp}$ can be used to distinguish decahedral and tetrahedral structures. Figures~2(c) and (d) show conformations which hardly differ because both of them possess 24 neighbor-neighbor contacts. They occur in almost all geometries considered here. Only the truncated fcc-octahedron (i.e., the ground state of the 38mer) does not exhibit triangular orthobicupolae [Fig.~\ref{fig:fig_2}(d)]. Nevertheless, since cuboctahedra [Fig.~\ref{fig:fig_2}(c)] are related to fcc- and triangular orthobicupolae to hcp-packing, their observation is of some use in other cases. In this study, we will mainly focus on the analysis of the number of complete and defected icosahedral cells, $n_{\rm ic}$.

\section{Simulation Method}
For the precise simulation of this model we have to face several challenges. Firstly, since the elastic bonds represent an additional restriction which causes multiple energy barriers, it is necessary to generate an appropriate simulation dynamics that allows the tunneling through these barriers and fast conformational changes in order to reduce autocorrelation times. We achieve the latter by applying three different conformational updates\revis{. The first causes a relatively small change by a simple shift of a monomer, the second propagates a monomer along the chain to a new position which is defined within a suitable local coordinate system, whereas the third alters only the linkage by swapping two bonds between four nearby monomers.} Details of these update procedures will be reported elsewhere \cite{cpc}.
\begin{figure}
\centerline{\epsfxsize=8cm \epsfbox{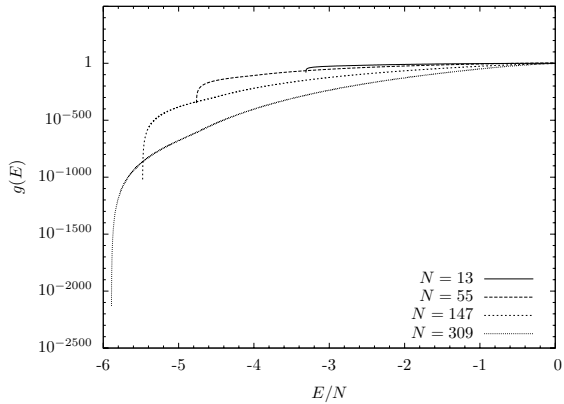}}
\caption{\label{fig:fig_3} Density of states for systems forming complete icosahedra.}
\end{figure}

Secondly, investigating the thermodynamic behavior of the polymer within the temperature interval of interest requires the sampling of regions of energy space with spectral densities differing by many hundreds to thousands orders of magnitude (Fig.~\ref{fig:fig_3}). In order to enable the system to perform a random walk in energy space we applied the multicanonical method \cite{muca}, where in iterative runs the multicanonical weight function $W(E)$ is estimated \cite{muca_it}. In a final production run \revis{a conformation $\mathbf{X}$ contributes with a generalized Boltzmann weight $e^{-\beta E(\mathbf{X})}W(E(\mathbf{X}))$ such that all energies are roughly equally probable. The density of states can then be determined by}
\begin{equation}
	g(E)\propto\frac{H(E)}{W(E)}e^{\beta E},
\end{equation}
where $H(E)$ is the histogram of the production run. Thereby, the Metropolis \cite{Metropolis} acceptance probability \revis{for a change from system conformation $\mathbf{X}$ to $\mathbf{X'}$ is replaced by
\begin{equation}
	P_{\rm accept}(\mathbf{X},\mathbf{X'})=\min\left(1,e^{-\beta (E(\mathbf{X'})-E(\mathbf{X}))}\frac{W(E(\mathbf{X'}))}{W(E(\mathbf{X}))}\right).
	\label{accept_prop}
\end{equation}}
The parameter $\beta$ influences the simulation only at the beginning of the weight estimation whereas it is canceled out when the weight function converges. To determine the multicanonical weights we performed several hundred single simulations, each with $10^5N$ updates. The final production run consisted of at least $10^9N$ updates. The obtained histogram can be reweighted to any desired temperature as long as the corresponding energy interval is covered by the simulation data, yielding the canonical energy distribution for this temperature.

Finally, the shape of the energy landscape might prevent the exploration of a narrow global energy minimum with standard Monte Carlo methods in certain cases -- a problem which is well-known from LJ clusters consisting of 38, 75-77, 98 or 102-104 atoms \cite{LJ98_102,LJ98_102long} -- so that further improvements of the simulation method are required. \revis{As described in the previous section, the number of icosahedral cells $n_{\rm ic}(\mathbf{X})$ allows for the geometrical classification of conformations $\mathbf{X}$. Based on $n_{\rm ic}(\mathbf{X})$, parameters $\nu(\mathbf{X})$ can be introduced to divide the conformational space in different parts. If non-icosahedral minima shall be explored, conformations without icosahedral cells must be treated differently from icosahedral conformations. Using
\begin{equation}
	\nu(\mathbf{X})=\left\{\begin{array}{l}
1\qquad {\rm if}\qquad n_{\rm ic}(\mathbf{X})=0\\
2\qquad {\rm if}\qquad n_{\rm ic}(\mathbf{X})\ge1,
\end{array} \right .
\end{equation}
the statistical weight of the two fractions can be enhanced or suppressed within the framework of multicanonical Monte Carlo sampling by employing multiple weight functions. Within an extended multicanonical ensemble a conformation  $\mathbf{X}$ is than represented with a probability
\begin{equation}
	p(\mathbf{X})\propto e^{-\beta E(\mathbf{X})}W_{\nu(\mathbf{X})}(E(\mathbf{X}))
\end{equation}
that depends on $E(\mathbf{X})$ and $\nu(\mathbf{X})$ such that (\ref{accept_prop}) changes to
\begin{eqnarray}
	P_{\rm accept}(\mathbf{X},\mathbf{X'})=\hspace{5cm}\nonumber\\
\min\left(1,e^{-\beta (E(\mathbf{X'})-E(\mathbf{X}))}\frac{W_{\nu(\mathbf{X'})}(E(\mathbf{X'}))}{W_{\nu(\mathbf{X})}(E(\mathbf{X}))}\right).
\end{eqnarray}}
\noindent Consequently, different geometries \revis{which correspond to different values of $\nu$} can be tuned to participate equally at any energy and free-energy valleys of different depths and widths can be sampled easily. \revis{Therefore,} the investigation of the solid-solid transitions is no longer a problem of barriers and huge autocorrelation times. While following the branches of different geometry, the only remaining task is to reach energies that correspond to a temperature which is sufficiently smaller than the temperature of the solid-solid transition.

\revis{In a similar way the system can be enabled to change more frequently between Mackay and anti-Mackay conformations. By using the parameter
\begin{equation}
	\nu'(\mathbf{X})=\left\{\begin{array}{l}
1\qquad {\rm if}\qquad n_{\rm ic}(\mathbf{X})\le1\\
2\qquad {\rm if}\qquad n_{\rm ic}(\mathbf{X})\ge2
\end{array} \right .
\end{equation}
and introducing corresponding weight functions, the simulations can be tuned to hit both types of structures with equal feqeuency at any energy as long as such conformations exist.}

%
\section{Discussion and Results}

\subsection{Ground states}
\begin{figure}
\centerline{\epsfxsize=8cm \epsfbox{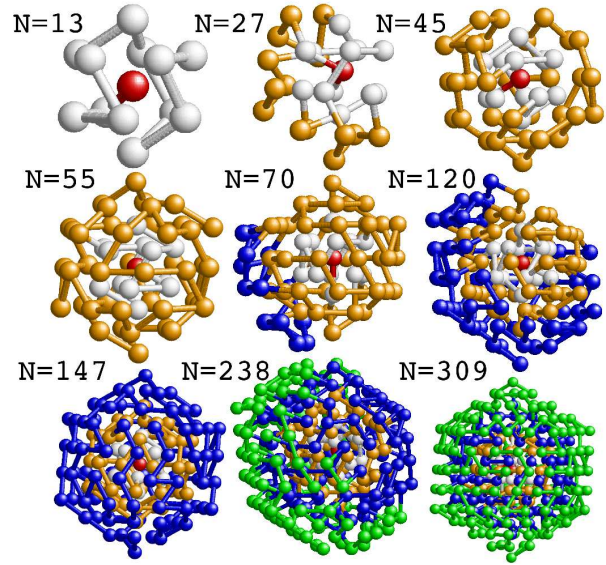}}
\caption{\label{fig:fig_4} (Color) Icosahedral (putative) ground-state conformations for different system sizes.}
\end{figure}

\begin{figure}
\centerline{\epsfxsize=8cm \epsfbox{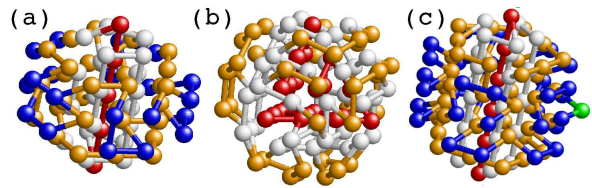}}
\caption{\label{fig:fig_5} (Color) Non-icosahedral ground-state conformations: (a) $N=75$, $r_c=2.5\sigma$ decahedral, (b) $N=98$, $r_c=5\sigma$ tetrahedral, (c) $N=102$, $r_c=5\sigma$ decahedral.}
\end{figure}

Global energy minimum conformations of the investigated polymers reveal wide similarities to ground-state configurations of LJ clusters, i.e., for almost all system sizes, the ground state is of icosahedral type (Fig.~\ref{fig:fig_4}). At zero temperature, the monomer-monomer bonds cause only small deviations since the minimum of the FENE potential is close to the equilibrium distance. Besides, the chain can arrange within the mostly unaltered monomer conformation to minimize the bond potential. As a result, bonds between different shells of an icosahedral core are rare since the corresponding monomer distances are smaller than the equilibrium distance and entail higher bond energies. For longer chains, the central icosahedral cell is strongly compressed and one end of the chain is usually located in the center of the cell, thus avoiding the inclusion of the most inappropriate distance (from the central monomer to one of its neighbors) in the chain twice. \revis{A similar effect occurs in decahedral conformations. Bonds between monomers on the central axis are favorable since their length is close to the optimum. This forces the polymer chain to adapt this axis at low temperatures [Fig.~\ref{fig:fig_5}(a),(c)].}

\begin{figure}
\centerline{\epsfxsize=6cm \epsfbox{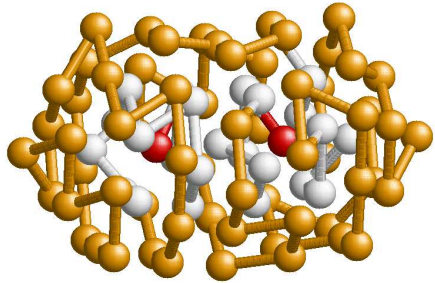}}
\caption{\label{fig:fig_6} (Color) The structure of minimal energy for $N=87$, $r_c=2.5\sigma$ is formed by two entangled icosahedral cores. The ends of the chain coincide with the respective centres.}
\end{figure}

We find perfect icosahedra for system sizes $N=13,55,147,309$. If the system size exceeds this 'magic' sizes, the polymer builds an icosahedral core with an anti-Mackay overlayer which grows with the chain length. At some point ($N>30$ and $N>80$), the overlayer adopts the structure of the core by changing to the Mackay type. Further increase of $N$ completes the outer shell and leads to the next icosahedron. A few polymers of certain sizes show ground-state conformations that correspond to different non-icosahedral geometries. We find a truncated octahedron for $N=38$ and a decahedral configuration for $N=75-77$ [Fig.~\ref{fig:fig_5}(a)]. Some deviations are caused by the cutoff of the LJ potential, so the chains with $N=81,85,87,98,102$ monomers possess lowest-energy structures that do not correspond to the respective clusters unless an untruncated LJ potential is applied. Using the cutoff $r_c=2.5\sigma$ we find for $N=81,85,98,102$ icosahedral ground states with Mackay overlayer and for $N=87$ a conformation with two merged icosahedral cores (Fig.~\ref{fig:fig_6}).

\subsection{Thermodynamics of complete icosahedra}
\begin{figure}
\centerline{\epsfxsize=8cm \epsfbox{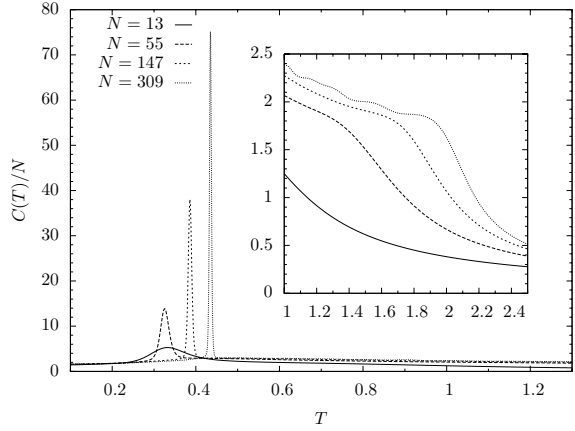}}
\caption{\label{fig:fig_7} The specific heats for chains forming complete icosahedra ($r_c=2.5\sigma$). Peaks indicate the melting transition at $0.3<T<0.5$, whereas the coil-globule collapse shown in the inset is less pronounced.}
\end{figure}

\begin{figure}
\centerline{\epsfxsize=8cm \epsfbox{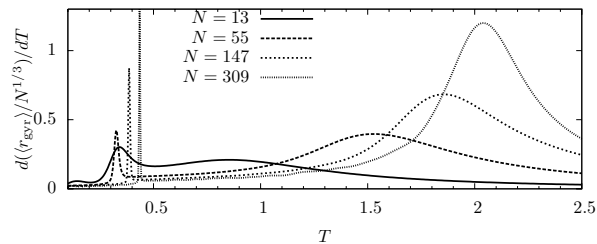}}
\caption{\label{fig:fig_8} The normalized fluctuations of radii of gyration for chains forming complete icosahedra give clear evidence for the freezing and the collapse transition as well as their increasing separation.}
\end{figure}

The four investigated chains that form complete icosahedra ($N=13,55,147,309$) exhibit a very clear and uniform thermodynamic behavior. We observe two separate conformational transitions indicated by peaks in the specific $C(T)$ heat and the fluctuations of the radius of gyration ${d\langle r_{\rm gyr}\rangle}/{dT}$. As shown in Figs.~\ref{fig:fig_7} and \ref{fig:fig_8}, the icosahedra melt in the interval $0.3<T<0.5$ and a liquid-like regime is reached where the polymer arranges still in a globular shape but exhibits no distinct structure. Hence, the icosahedral order parameter $\langle n_{\rm ic}\rangle \approx 1$ changes to $\langle n_{\rm ic}\rangle \approx 0$. The corresponding peak in the normalized specific heat increases rapidly with system size and allows in principle a precise determination of the melting temperature. However, all solid-solid or liquid-solid transitions considered here must not be understood as thermodynamic transitions in a strict sense since all investigated systems are small and dominated by finite-size effects. For longer chains one would expect, in analogy to the thermodynamic behavior of LJ clusters, the crossover from icosahedral ground states to decahedral ($N\gtrsim1500$) and later to fcc-like \cite{DoyeCalvo} structures ($N\gtrsim20000$) which exhibit a different crystallization behavior. Therefore the extrapolation to the thermodynamic limit by means of finite-size scaling is unfortunately not an appropriate choice. It is worth noting that for $N=309$ the result differs considerably from the pure LJ cluster in which case a recent study \cite{LJcluster_309} found two clearly separated peaks.

Increasing $T$ further leads to the collapse transition at temperatures $1\le T\le 2$. Above this temperature the chains arrange randomly in extended conformations. For almost all investigated systems this cross-over does not produce a clear signal in the specific heat except for the chain with $N=309$ where a shoulder emerges at $T\approx1.8$. If the chain is shorter this energetic signal of the collapse transition is suppressed by finite-size effects. However, geometric quantities like the radius of gyration and its fluctuation (Fig.~\ref{fig:fig_8}) give some insight. It is obvious that the solid-liquid transition remains well separated from the coil-globule collapse, moreover the intermediate temperature interval increases within the investigated chain length interval.

The solid phase is dominated by the extremely stable icosahedron which, however, depending on the temperature, exhibits surface defects. Although the mobility of a great majority of the monomers is strongly restricted, there are  still changes which can be observed. Firstly, the chain, i.e., the linkage of the monomers still transforms relatively freely. Only at extreme low temperatures the lowest energy (chain-) conformation gains thermodynamic relevance. Starting from the center, the number of bonds connecting different shells is reduced. At $T=0$ the first three shells are connected only by a single bond, so that for $N=13,55,147$ one end of the chain is the central monomer and the other is at the surface. For $N=309$ bonds between the corners of the third and the fourth shell exist also at $T=0$, since the length of this bonds is roughly equal to alternative bonds within the third shell, which include the corner monomers. Secondly the entire icosahedron undergoes a compactification with decreasing temperature.

\subsection{Liquid-solid transitions of elastic polymers}

In the following, we analyze the behavior of small elastic polymers in the liquid-solid transition regime. Particular emphasis will be dedicated to the chain-length dependence of geometric changes in the conformations of the polymers while passing the transition line.

\begin{figure}
\centerline{\epsfxsize=8cm \epsfbox{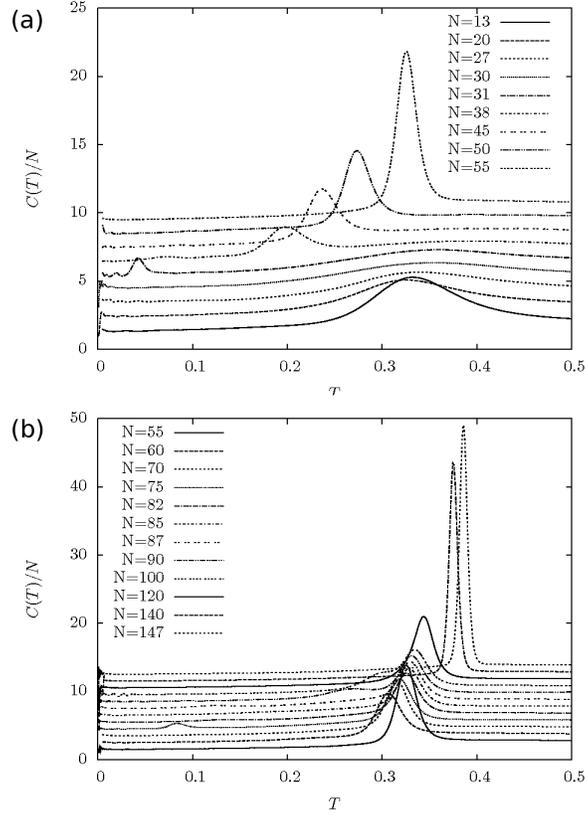}}
\caption{\label{fig:fig_9} Specific heat for (a) $13\le N\le 55$ and (b) $55\le N\le 147$, $r_c=2.5\sigma$. For better visibility, the curves are shifted by a constant offset.}
\end{figure}

The size dependence of the specific heat for polymers with sizes up to $N=55$ is illustrated in Fig.~\ref{fig:fig_9}(a). The ground state of the short 13mer is the energetically stable icosahedron whose almost perfect symmetry is only slightly disturbed by the FENE bond potential. The melting transition is indicated by a the peak at $T=0.33$. With increasing system size the low-temperature conformations are less symmetric and the peak becomes broader. For $N>30$, the crossover from Mackay-like ground states to anti-Mackay conformations takes place, the corresponding peak increases with growing $N$ and shifts from $T\approx0.042$ to higher temperatures. Finally, for $N=55$, there is only one melting transition left and anti-Mackay-like structures are strongly suppressed. As an effect of the bond elasticity, the melting transitions occur at slightly higher temperatures than in the case of pure LJ clusters \cite{LJcluster_heatcap}.

\begin{figure}
\centerline{\epsfxsize=8cm \epsfbox{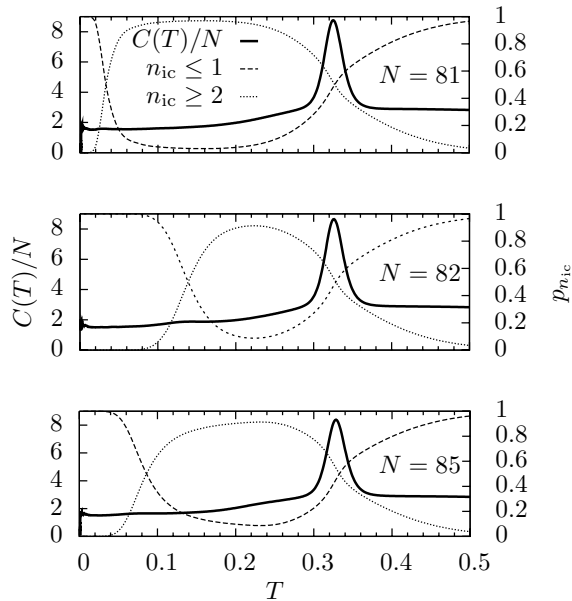}}
\caption{\label{fig:fig_10} Whereas the Mackay--anti-Mackay crossover is not recognizable in the specific heat curves, the probability of occurrence of specified numbers of icosahedral cells, $p_{n_{\rm ic}}$, reveals the transition temperature ($r_c=2.5\sigma$).}
\end{figure}

\begin{figure}
\centerline{\epsfxsize=8cm \epsfbox{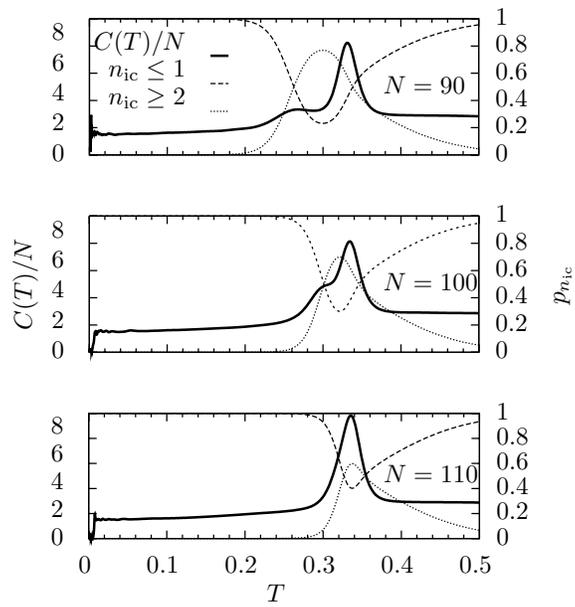}}
\caption{\label{fig:fig_11} Already for $N\ge110$ the entire solid phase is dominated by Mackay ($n_{\rm ic}=1$) conformations ($r_c=2.5\sigma$).}
\end{figure}

The behavior of chains containing between 55 and 147 monomers [Fig.~\ref{fig:fig_9}(b)] again generally corresponds to that of LJ clusters \cite{LJcluster_heatcap} with the differences that the melting transition occurs at higher temperatures and that the Mackay--anti-Mackay transition temperature increases much faster with system size. Besides there are a few chains with ground states of types different to LJ clusters which is induced by the truncation of the LJ potential as mentioned above. We find anti-Mackay ground states up to system sizes of $N=80$. In contrast to the clusters, $N=81$ and $N=85$ possess a GEM of Mackay type. Hence we encounter a solid-solid transition also in these cases. We do not find peaks of the solid-solid transition in the specific heat for $N=81,82,85$, and the crossover to anti-Mackay conformations can only be identified in changes of \revis{the structural ``order'' parameter $n_{\rm ic}$ defined in (\ref{p_n_ic}). In Fig.~\ref{fig:fig_10} we therefore also show the probability $p_{n_{\rm ic}}$ that $n_{\rm ic}\le 1$ or $n_{\rm ic}\ge2$ as a function of temperature.} While one would expect higher transition temperatures for growing system size, this prediction fails in the case of $N=85$, where the anti-Mackay energy minimum is almost as deep as the GEM due to an optimal arrangement of the outer monomers. Nevertheless, for $N\ge90$, the transition shifts rapidly to higher temperatures (Fig.~\ref{fig:fig_11}) and manifests in the specific heat as well. Whereas in the case of pure LJ clusters, the two peaks remain separated up to sizes of 130 atoms \cite{LJcluster_heatcap}, we observe both transitions merging already for $N\approx 100$. In contrast to the polymers, the Mackay--anti-Mackay transition temperature of clusters even decreases near $N=120$, presumably because the anti-Mackay overlayer is almost complete, leading to a spherical and therefore stable conformation. This indicates that in this polymer model anti-Mackay structures loose weight compared to atomic clusters.
For $N=75$, one finds the crossover between decahedral and icosahedral conformations indicated by a small peak at $T\approx 0.8$.

\subsection{The cutoff's influence}
\begin{figure}
\centerline{\epsfxsize=6cm \epsfbox{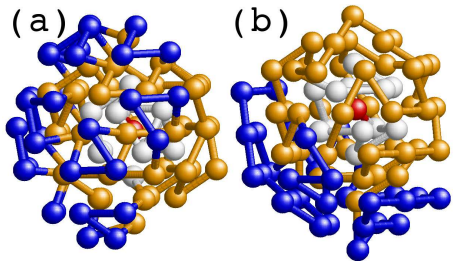}}
\caption{\label{fig:fig_12} (Color) Conformations of minimal energy for $N=85$ with (a) anti-Mackay and (b) Mackay overlayer.}
\end{figure}

\begin{figure}
\centerline{\epsfxsize=8cm \epsfbox{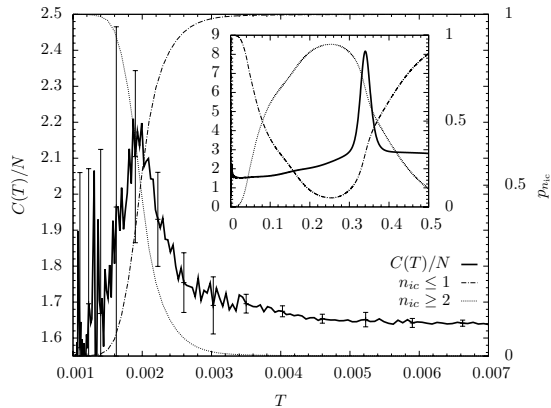}}
\caption{\label{fig:fig_13} The polymer with $N=85$ and untruncated LJ potential ($r_c=\infty$) changes from anti-Mackay ($T<0.002$) to Mackay conformations ($0.002<T<0.1$) and back ($0.1<T<0.34$).}
\end{figure}

At this point we cannot yet judge whether the differences to the behavior of atomic clusters at medium temperatures are caused by the truncation of the LJ potential or by the polymer topology, i.e., the bonds. To answer this question, we will discuss three interesting cases with modified cutoff. First, we consider the polymer $N=85$ with the original LJ potential, i.e., $r_c=\infty$. The corresponding cluster is the largest in the interval $55<N<147$ with an anti-Mackay ground state \cite{Northby}. Since we use the unaltered LJ potential, any deviation to the cluster behavior is caused by the bond potential only. In contrast to the previous simulations with truncated potential, we retain the anti-Mackay ground state where the 30 outer monomers completely cover 10 faces of the icosahedral core and build an energetically favored structure [Fig.~\ref{fig:fig_12}(a)]. One might notice that there are no bonds connecting monomers on different faces directly since the bond length would be too far from the potential minimum. This means that for very low energies only a few bond configurations are allowed and that the anti-Mackay state is much less metastable than the Mackay state [Fig.~\ref{fig:fig_12}(b)] for which many more bond configurations with low energies are possible. This leads to an entropic dominance of the latter in the temperature interval $0.002<T<0.08$ as visible in Fig.~\ref{fig:fig_13}, a behavior that has never been reported for atomic clusters. In spite of large statistical errors we find a signal in the specific heat at $T\approx 0.002$, since at this temperature the energetically favored anti-Mackay state prevails. The transition back to anti-Mackay conformations extends over a wider temperature interval around $T\approx 0.08$ and cannot be localized in the specific heat. Again the change from a LJ cluster to a polymer leads to a greater prominence of Mackay conformations.

\begin{figure}
\centerline{\epsfxsize=8cm \epsfbox{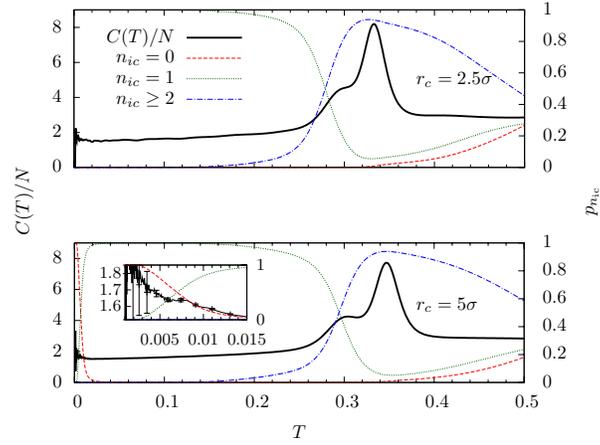}}
\caption{\label{fig:fig_14}\revis{(Color)}  Although there is almost no change in the specific heat the values $p_{n_{\rm ic}}$ indicate clearly the crossover to a non-icosahedral ground state if the cutoff $r_c$ is increased ($N=98$).}
\end{figure}

\begin{figure}
\centerline{\epsfxsize=8cm \epsfbox{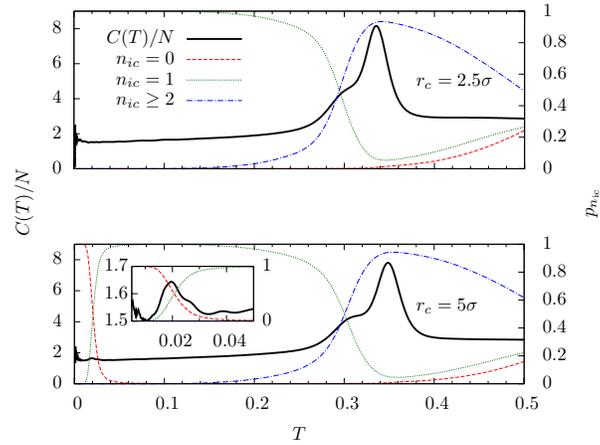}}
\caption{\label{fig:fig_15}\revis{(Color)}  Same as Fig.~\ref{fig:fig_14} for $N=102$. The solid-solid transition to the decahedral ground state leaves a small but distinct peak in the specific heat.}
\end{figure}

To prove the influence of the cutoff on the thermodynamics at higher temperatures we doubled the cutoff to $r_c=5\sigma\approx3.12$, minimizing differences to the untruncated potential and investigated the two systems $N=98,102$. For the corresponding clusters, two separate transitions where observed at medium temperatures which stands in contrast to the polymer with original cutoff ($r_c=2.5\sigma$) where both transitions merge. Applying the enlarged cutoff we observe only a slight shift in the peak positions whereas the major differences to the LJ clusters' behavior as demonstrated in Ref.~\cite{LJcluster_heatcap} persist (Figs.~\ref{fig:fig_14}, \ref{fig:fig_15}). It is still impossible to distinguish both conformational transitions clearly by means of the specific heat, i.e., the temperature domain where anti-Mackay like conformations prevail is very small. We may conclude that this difference to cluster behavior (i.e., the suppression of anti-Mackay states) is mainly an effect of the bonds.

The second interesting outcome of the simulations are the different ground states. As in the case of unbounded clusters with untruncated interaction, we obtain a GEM of tetrahedral geometry for $N=98$ [Fig.~\ref{fig:fig_5}(b)] and a decahedral ground state for $N=102$ [Fig.~\ref{fig:fig_5}(c)]. We can also identify the solid-solid transition to Mackay structures: For $N=98$, $\langle n_{\rm ic} \rangle \approx 0$ changes as expected to $\langle n_{\rm ic} \rangle \approx 1$ at $T\approx0.0075$ \revis{(see inset of Fig.~\ref{fig:fig_14})} and a tiny bulge is formed which, however, has to be considered with caution, since statistical errors increase strongly at such low temperatures. In the other case ($N=102$) the decahedral-icosahedral crossover occurs at a higher temperature $T\approx0.02$ (see Fig.~\ref{fig:fig_15}), with a more prominent signal in the specific heat.

After all we see some evidence that during the change from atomic LJ clusters to polymers, Mackay-like structures are more favored independently of a truncation of the LJ potential. This results in a shift of the Mackay--anti-Mackay (or Mackay overlayer melting-) transition to higher temperatures and the formation of a Mackay dominated temperature interval for the 85mer if the original LJ potential is employed.

\begin{figure}
\centerline{\epsfxsize=8cm \epsfbox{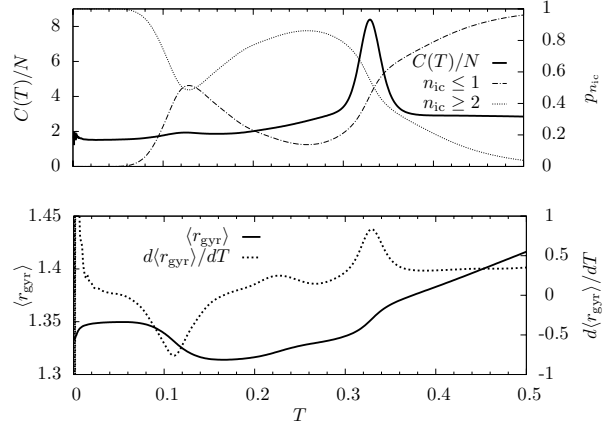}}
\caption{\label{fig:fig_16}For $N=87$ the transition to a unique GEM of excentric shape is visible in the specific heat, in $p_{n_{\rm ic}}$ as well as in the radius of gyration $r_{\rm gyr}$ which strongly increases during the transition.}
\end{figure}

A further clear effect of the truncation of the potential arises in the case of $N=87$. This particular number of monomers allows the formation of two merged icosahedral cores, each with a icosahedral center of high density (Fig.~\ref{fig:fig_6}). The overall shape of this conformation reminds of a cylinder and is therefore not optimal for the original long-range LJ potential which tends to form spherical structures. Applying the cutoff $r_c=2.5\sigma$, the short-range interactions gain importance and the double-core conformation appears to be the ground state. At $T\approx 0.12$ the system undergoes the cross-over to anti-Mackay structures and also Mackay conformations play a minor role (Fig.~\ref{fig:fig_16}). Since thereby the polymer becomes more spherical, we observe a decrease of the radius of gyration. Note that in the ground-state conformation (Fig.~\ref{fig:fig_6}) both ends of the chain are located in the centers of the two cores, thereby minimizing the chain energy in the way mentioned above. 
\begin{figure}
\centerline{\epsfxsize=8cm \epsfbox{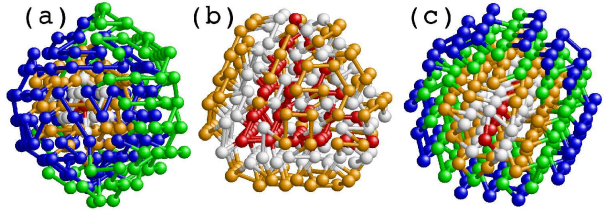}}
\caption{\label{fig:fig_17} (Color) Low-energy conformations of the polymer with $N=234$ ($r_c=5\sigma$) with (a) icosahedral, (b) tetrahedral, and (c) decahedral geometry.}
\end{figure}
\subsection{Non-icosahedral minima of larger systems}
To demonstrate the capability of our method to find different structures of minimal energy we now present results for a relatively long chain. Using a $5\sigma$-cutoff, we investigated the 234mer which has the suitable length to form tetrahedral as well as decahedral minima. To distinguish the two cases, we used the number of elongated pentagonal pyramids [Fig.~\ref{fig:fig_2}(b)] $n_{\rm epp}$ in addition to $n_{\rm ic}$ for the classification. While $n_{\rm epp} \le 6$ for conformations with decahedral symmetry, $n_{\rm epp} = 24$ in the tetrahedral case ($n_{\rm ic}=0$ for both). We were able to find the expected structures for each geometry (Fig.~\ref{fig:fig_17}) and a final minimization with the conjugate gradient technique \cite{con_grad} reveals that while the icosahedral minimum represents the GEM the energy difference to the minima of the two other geometries is very small (see Table \ref{tab:234all_min}). This is also the case when considering the minimized respective cluster conformations, i.e., neglecting bonds and applying the untruncated LJ potential. The obtained value for the icosahedral LJ cluster minimum is in perfect agreement with an earlier study \cite{Barron}, hence we may assume that also the energies of the decahedral and the tetrahedral state are close to the true minimal values.

\begin{table}[ht]
\caption{\label{tab:234all_min} Minimal energy values for conformations of different geometry of the polymer with $N=234$ ($r_c=5\sigma$) and the 234-atom LJ cluster.}
\begin{tabular}{cp{3mm}ccc}\hline\hline
   & & icosahedral & tetrahedral & decahedral\\ \hline
polymer & & $-1458.877$ & $-1458.431$ & $-1458.036$\\ 
\parbox{3cm}{LJ cluster} & & $-1465.924$ & $-1465.529$ & $-1465.116$\\ \hline \hline
\end{tabular}
\end{table}

\section{summary}
In this study, we demonstrated the capability of the multicanonical sampling method to investigate the complete thermodynamic behavior of flexible elastic polymer chains containing more than 300 monomers within reasonable time on a standard workstation. In order to achieve this we developed new conformational updates which enhanced the performance significantly. Analyzing the behavior of the radii of gyration it became evident that the solid-liquid transition and the coil-globule collapse will remain well separated also for much longer chains. In our study of conformational properties we focused on the low-temperature regime. It turned out that the extensive results of the last decades' research on the subject of atomic Lennard-Jones clusters provide an excellent framework for also understanding the liquid-solid transition behavior of the polymers. To observe the expected low-temperature transitions including conformations of completely different geometry, we introduced the total number of icosahedral cells $n_{\rm ic}$ as an order parameter for the classification and distinction of the respective structures. Fortunately, this new quantity did not only prove useful in visualizing the conformational changes, but also enabled us to explore the tetrahedral and decahedral minima for $N=75,98,102$, and $234$. For this purpose we introduced multiple weight functions which is a relative simple modification to the standard multicanonical method which should be in principle also adaptable to other methods like parallel tempering. 

The low-temperature behavior of elastic polymers exhibits strong similarities to atomic Lennard-Jones clusters. In almost all cases, the global energy minimum possesses an icosahedral core and an overlayer of anti-Mackay (hcp) or Mackay (fcc) type. Differences in the putative ground states (e.g., for $N=81,87,98,102$) could be ascribed to the truncation of the Lennard-Jones potential. Furthermore, in our analysis of the long-range cutoff influence on crystallization we made manifest for $N=87$ and $r_c=2.5\sigma$ the solid-solid transition from icosahedral double-core to single-core conformations and for $N=75$, $r_c=2.5\sigma$ and $N=98,102$, $r_c=5\sigma$ the cross-over from non-icosahedral to icosahedral structures. As a common effect of the bonds, all melting transitions were shifted to higher temperatures compared to atomic Lennard-Jones clusters. Besides, for chains of length  $55\le N\le 147$ further deviations arose: While in the case of clusters for $82\le N\le 130$ the Mackay--anti-Mackay transition was clearly separated from the core-melting this only holds for polymer chains with $81\le N<100$. For longer chains both transitions merge. For the chain with $N=85$ and untruncated Lennard-Jones potential, we find a temperature range where Mackay conformations dominate, embedded in regions of anti-Mackay structures. We regard this last two findings as evidence for the tendency of the bond potential to suppress anti-Mackay structures.

\section*{Acknowledgements}
This work is partially supported by the DFG (German Science Foundation) under Grant  
Nos.\ JA 483/24-1/2/3 and the Leipzig Graduate School of Excellence ``BuildMoNa''. Support by a supercomputer time grant (hlz11) of the John von Neumann Institute for Computing (NIC), Forschungszentrum J\"ulich, is gratefully acknowledged. 
\pagebreak

\end{document}